\documentclass[aps,pre,epsf,twocolumn]{revtex4}
\usepackage{amsmath}
\usepackage{amssymb}
\usepackage{epsfig}
\usepackage{graphicx}
\usepackage[T1]{fontenc}
\usepackage[utf8]{inputenc}
\usepackage{color}
\usepackage[normalem]{ulem}
\usepackage{tikz}
\usetikzlibrary{positioning}

\graphicspath {{./figures/}}
\makeatletter
\def\input@path{{./figures/}}
\makeatother

\begin{document}
\title{Universality in the two-dimensional dilute Baxter-Wu model}

\author{Alexandros Vasilopoulos$^1$}

\author{Nikolaos G. Fytas$^1$}
\altaffiliation[]{Corresponding author: nikolaos.fytas@coventry.ac.uk}

\author{Erol Vatansever$^{2,1}$}

\author{Anastasios Malakis$^{3,1}$}

\author{Martin Weigel$^{4}$}

\affiliation{$^1$Centre for Fluid and Complex Systems, Coventry
	University, Coventry, CV1 5FB, United Kingdom}

\affiliation{$^2$Department of Physics, Dokuz Eyl\"{u}l University, TR-35160, Izmir, Turkey}

\affiliation{$^3$Department of Physics, University of Athens, Panepistimiopolis, GR 15784 Zografou, Greece}

\affiliation{$^4$Institut für Physik, Technische Universität Chemnitz, 09107 Chemnitz, Germany}

\date{\today}

\begin{abstract}
We study the question of universality in the two-dimensional spin-$1$ Baxter-Wu model in the presence of a crystal field $\Delta$. We employ extensive numerical simulations of two types, providing us with complementary results: Wang-Landau sampling at fixed values of $\Delta$ and a parallelized variant of the multicanonical approach performed at constant temperature $T$. A detailed finite-size scaling analysis in the regime of second-order phase transitions in the $(\Delta, T)$ phase diagram indicates that the transition belongs to the universality class of the $4$-state Potts model. Previous controversies with respect to the nature of the transition are discussed and possibly attributed to the presence of strong finite-size effects, especially as one approaches the pentacritical point of the model.
\end{abstract}

\maketitle

\section{Introduction}
\label{sec:intro}

\begin{figure}
	\includegraphics[width=70mm]{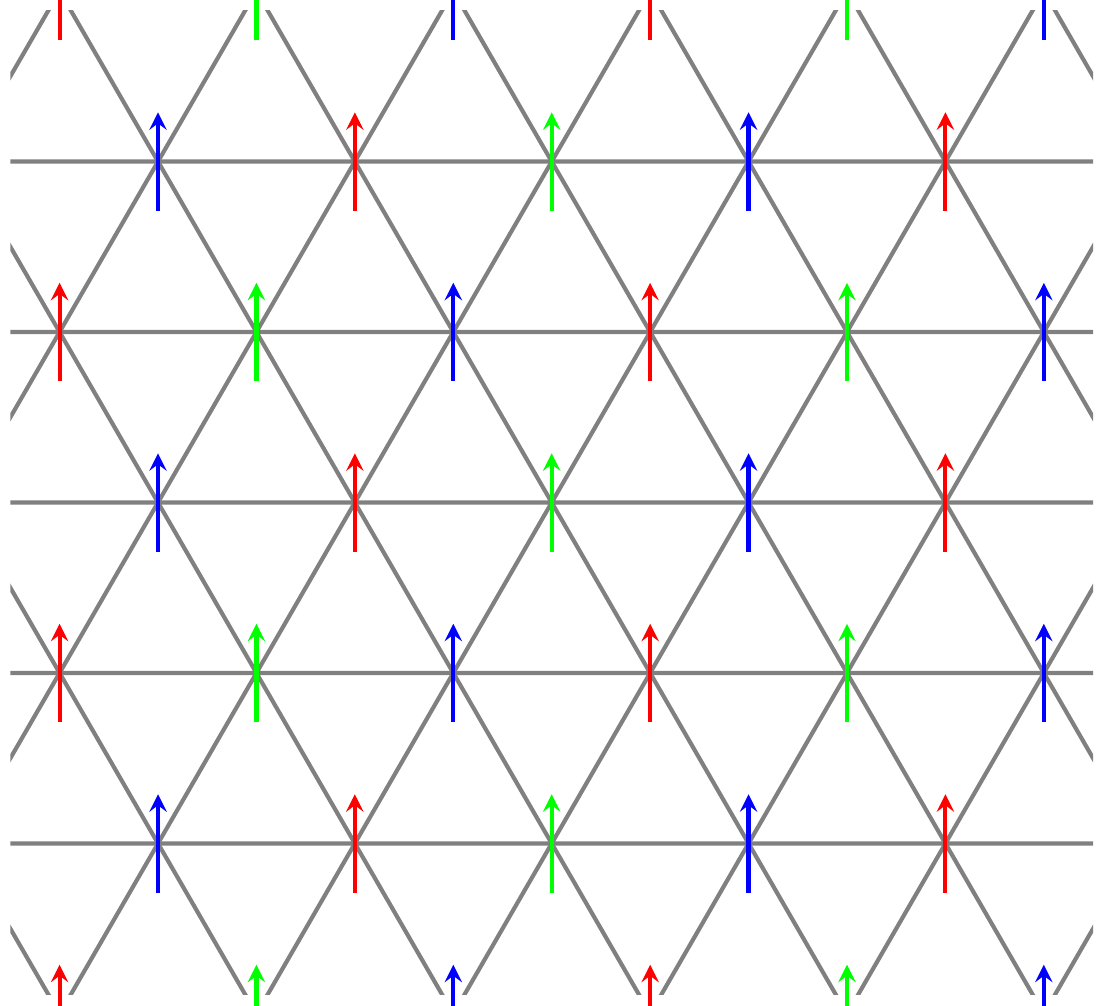}
	\caption{Representation of the Baxter-Wu triangular lattice as a superposition of the three sublattices A, B, and C. Each sublattice corresponds to spins of different color. The spins are shown in the ferromagnetic ground state.}  
		\label{fig:lattice}
\end{figure}

The Baxter-Wu (BW) model was first introduced by Wood and Griffiths~\cite{wood72} as a system which does not exhibit invariance under a global inversion of all spins. It is defined on a triangular lattice by the Hamiltonian
\begin{equation}\label{eq:Ham}
  \mathcal{H}_{\rm BW}
  = -J\sum_{\langle xyz \rangle}\sigma_{x}\sigma_{y}\sigma_{z},
\end{equation}
where the exchange interaction $J$ is positive, the sum extends over
all elementary triangles of a lattice with $N$ sites,
and $\sigma_{x} = \pm 1$ are Ising spin-$1/2$ variables. The triangular lattice can be divided into three sublattices A, B, and C as shown in Fig.~\ref{fig:lattice}, so that any triangular face contains one site of type A, one of type B, and one of type C. The ground state of the model is four-fold degenerate: there is one ferromagnetic state with all spins up, and three ferrimagnetic states with down spins in two sublattices and up spins in the third sublattice. Also, the model of Eq.~(\ref{eq:Ham}) is self-dual~\cite{wood72,merlini72}, having the same critical temperature
as the spin-$1/2$ Ising model on the square lattice, i.e.,
$k_{\rm B}T_{\rm c}/J =2/\ln{(\sqrt{2}+1)} = 2.269185\ldots$, where $k_{\rm B}$ denotes the Boltzmann constant. 

The exact solution of Baxter and Wu dates back to 1973 and provided the critical exponents $\alpha = 2/3$, $\nu = 2/3$, and $\gamma = 7/6$~\cite{baxter73,baxter_book}. In the following, it was shown that the critical behavior of the model corresponds to a conformal field theory with central charge $c = 1$~\cite{alcaraz97,alcaraz99}. As was first pointed out by  Domany and Riedel, the $q = 4$ Potts model should belong to the same universality class as the Baxter-Wu model, as both have the same symmetry and degree of degeneracy in the ground state~\cite{domany78}. However, although the leading critical exponents are the same, one should note that these two models have different corrections to scaling: while the $4$-state Potts model presents logarithmic corrections with the system size, as expected for the marginal case before the transition becomes first-order for $q > 4$~\cite{wu82}, the Baxter-Wu model has power-law corrections with a correction-to-scaling exponent $\omega = 2$~\cite{alcaraz97,alcaraz99}. This rather large value of $\omega$ allows for a safe determination of the asymptotic scaling behavior even when dealing with systems of moderate size, see for instance Ref.~\cite{jorge16}. Recently, further aspects of the spin-$1/2$ model have also been considered, including short-time dynamics~\cite{hadjiagapiou05}, Monte Carlo studies of critical amplitude ratios~\cite{shchur10}, longitudinal~\cite{velonakis13}, and transverse~\cite{capponi14} magnetic fields.

An interesting extension of the Baxter-Wu model~(\ref{eq:Ham}) arises when one considers spin values $\sigma_{x} =  \{-1,0,1\}$ and includes an extra crystal field (or single-ion anisotropy) $\Delta$, so that the resulting Hamiltonian reads
\begin{equation}\label{eq:Ham2}
  \mathcal{H}
  = -J\sum_{\langle xyz \rangle}\sigma_{x}\sigma_{y}\sigma_{z}+\Delta\sum_{x}\sigma_{x}^{2} = E_{J}+\Delta E_{\Delta}.
\end{equation}
In the following we will use reduced units where $J=1$ as well as $k_B = 1$.
Unfortunately, for this model there exists no exact solution and therefore approximation methods need to be employed. Note, however, that when $\Delta \rightarrow -\infty$ only configurations with $\sigma_{x} = \pm 1$ are allowed and the pure Baxter-Wu model is recovered.

\begin{figure}
\includegraphics[width=85mm]{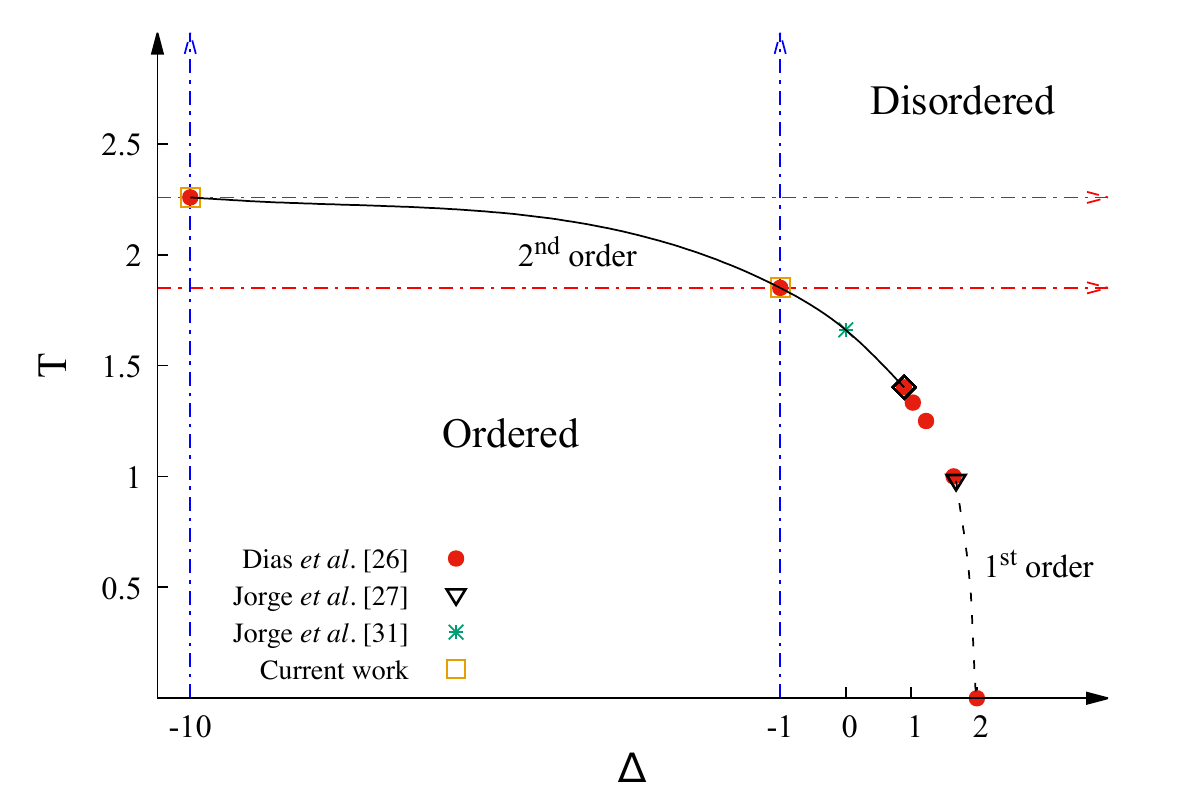}
\caption{Phase diagram of the two-dimensional spin-$1$ Baxter-Wu model. Several transition points are given including those obtained in the current work. The black rhombus and black triangle mark the pentacritical point as estimated by Dias \emph{et al.}~\cite{dias17} $(\Delta_{\rm pp}, T_{\rm pp}) \approx (0.8902, 1.4)$, and Jorge \emph{et al.}~\cite{jorge21} $(\Delta_{\rm pp}, T_{\rm pp}) \approx (1.68288(62), 0.98030(10))$, respectively. The black dotted and continuous lines correspond to first- and second-order phase transitions. The intermediate regime between the two pentacritical point estimations is not crossed by a line as it calls for further investigation. Blue vertical and red horizontal dotted arrows indicate the two numerical approaches used, namely Wang-Landau and multicanonical methods at fixed values of $\Delta = \{-10,\;-1\}$ and $T=\{2.2578, \;1.8503\}$, respectively.
\label{fig:phase_diagram}}
\end{figure}

As is apparent, the model of Eq.~(\ref{eq:Ham2}) resembles the well-known Blume-Capel model~\cite{blume66}, which exhibits a phase diagram with ordered ferromagnetic and disordered paramagnetic phases separated by a transition line with first- and second-order segments (the latter in the Ising universality class) connected by a tricritical point. More details about the phase diagram and universality aspects of the general Blume-Capel model can be found in Refs.~\cite{oliveira95,xavier98,plascak03,malakis09,malakis10,zierenberg15,kwak15,zierenberg17}.
In analogy to these findings, one might expect for the model defined in Eq.~(\ref{eq:Ham2}) a similar phase diagram but a different universality class. Nienhuis \emph{et al.}~\cite{nienhuis79} first discussed the analogy between the Baxter-Wu and diluted Potts models and pointed out that the general phase diagram will exhibit a line of continuous transitions that connects to a regime of first-order transitions through a multicritical point. Kinzel \emph{et al.}~\cite{kinzel81}, instead, using a finite-size scaling method, conjectured that a continuous transition only occurs for  $\Delta \rightarrow -\infty$ (the pure Baxter-Wu model). More recent work has favored the existence of a multicritical point at finite values of $\Delta$~\cite{costa04}. In Ref.~\cite{dias17} the location of the pentacritical point was estimated as $(\Delta_{\rm pp}, T_{\rm pp}) \approx (0.8902, 1.4)$, whereas Jorge \emph{et al.}~\cite{jorge21} suggested $(\Delta_{\rm pp}, T_{\rm pp}) \approx (1.68288(62), 0.98030(10))$, see Fig.~\ref{fig:phase_diagram} but also Fig. 5 of Ref.~\cite{jorge21} for a reproduction of the phase diagram of the model. This pentacritical point refers to the coexistence of three ferrimagnetic configurations and a ferromagnetic configuration, along with that of zero spins. The results of Ref.~\cite{dias17} for the critical exponents $\nu \approx 0.63$ and $\eta \approx 0.23$ point to the universality class of the pure spin-$1/2$ Baxter-Wu model where $\nu = 2/3$ and $\eta = 1/4$. 

Surprisingly though, there are still open questions with respect to the universality principle of the spin-$1$ Baxter-Wu model. The results of Ref.~\cite{costa04} via renormalization group, conventional finite-size scaling, and conformal invariance techniques indicated that the critical exponents vary continuously with $\Delta$ along the second-order transition line, differently from the expected behavior of the $4$-state Potts model. A similar conclusion was drawn in Ref.~\cite{costa04b}, where using importance sampling Monte Carlo simulations for the special case with $\Delta = 0$ the values $\nu = 0.617(3)$, $\alpha = 0.692(6)$, and $\gamma  = 1.13(1)$ were obtained. The complementary Monte Carlo results of Ref.~\cite{costa16} for $\Delta = -1$ and $1$ further corroborated this hypothesis~\cite{comment1}. Conversely, the renormalization-group work of Dias \emph{et al.}~\cite{dias17} suggested that along the critical line, the conformal anomaly $c$ and the exponents $\nu$ and $\eta$ are the same as that of the pure spin-$1/2$ Baxter-Wu model (or the $4$-state Potts model). The most recent work by Jorge \emph{et al.}~\cite{jorge20} used Wang-Landau sampling to probe the system's behavior at $\Delta = 0$. According to these authors it exhibits an indeterminacy regarding the order of phase transition, \emph{i.e.}, the analysis of numerical data was conclusive for both types of transitions, continuous or of first-order type. For the former case they estimated the values $\nu = 0.6438(10)$ and $\gamma = 1.1521(13)$. Finally, recent numerical evidence at the first-order transition regime of the phase boundary suggested that the specific heat exhibits a double peak structure as in the Schottky-like anomaly, which is associated with an order–disorder transition~\cite{jorge21}.

In the present work we provide a resolution of these controversies. Using extensive numerical simulations, as outlined in Sec.~\ref{sec:methods} below, we scrutinize the critical properties of the model, covering the whole extent of the continuous transition line. In particular, in an attempt to identify the presence and role of finite-size effects, we perform Wang-Landau simulations at two fixed values of the crystal field, $\Delta = -10$, deep in the second-order regime, and $\Delta = -1$ in the vicinity of the pentacritical point. We complement these by multicanonical simulations at the temperature $T = 1.8503$ crossing the phase boundary at $\Delta \approx -1$ as indicated in Fig.~\ref{fig:phase_diagram}. The remainder of the paper is organized as follows: In Sec.~\ref{sec:methods} we outline the Wang-Landau and parallel multicanonical simulation methods that we use to study the problem, and we introduce the observables studied. Our numerical results and the relevant finite-size scaling analysis are presented in Sec.~\ref{sec:results}. Finally, in Sec.~\ref{sec:summary} we summarize our findings and provide an outlook.

\section{Numerical Methods and Observables}
\label{sec:methods}

We use a combination of Wang-Landau and multicanonical simulations in a complementary strategy. This combined scheme allows us to cross the phase boundary of the model in two directions (see the dotted arrows in Fig.~\ref{fig:phase_diagram}) and probe efficiently the critical properties of the model.

\begin{figure}
	\includegraphics[width=95mm]{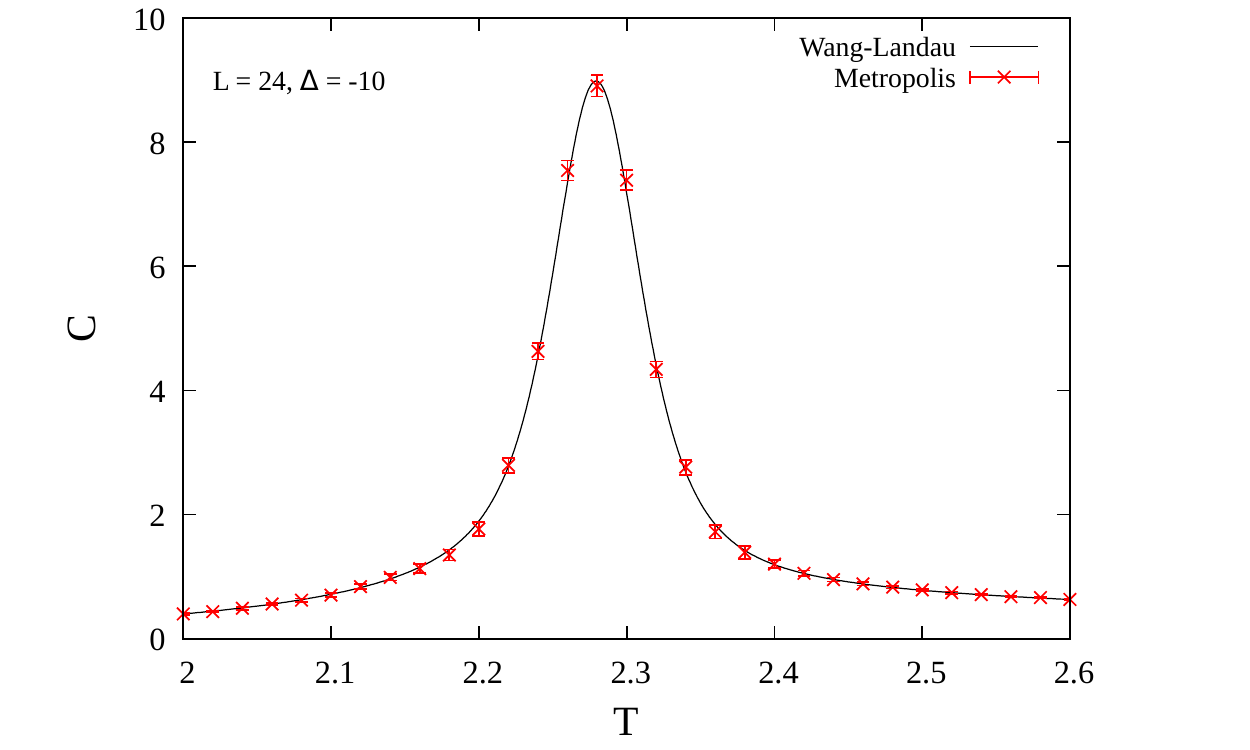}
	\caption{Specific-heat curves of the spin-$1$ Baxter-Wu model at $\Delta = -10$ for a system with linear size $L = 24$ obtained via Wang-Landau and Metropolis simulations.
		\label{fig:test}}
\end{figure}

\subsection{Wang-Landau simulations}
\label{sec:WL}

In a Wang-Landau simulation~\cite{wang01} random walks are performed in energy space and trial spin configurations are accepted with a probability proportional to the reciprocal estimate of the density of states, $1/g(E)$. The estimate $g(E)$ for the current energy is modified as $g(E) \rightarrow f \cdot g(E)$, where $f$ is known as the modification factor. During the simulation, an energy histogram is also accumulated. If this is flat, the modification factor is adjusted according to the rule $f_{j+1} = \sqrt{f_{j}}$, where $f_{1} = e$. In the present work we used a flatness criterion of $90\%$, as well as $j_{\rm final} = 24$. Furthermore, to increase statistical accuracy we averaged over several independent samples, typically $\sim 32$.

Our strategy follows the more stringent one-range implementation of the Wang-Landau algorithm, compared to the more efficient multi-range approach where one splits the energy range in many sub-intervals and joins the densities of states from the separate pieces at the end. This multi-range approach is almost a necessity for very large lattices and in many cases has produced results of high accuracy~\cite{wang01}. However, there are many subtleties with respect to boundary effects~\cite{netto08} and especially in cases where first-order transition characteristics appear~\cite{fytas08a}, hence justifying our choice. The simulations were facilitated by the use of restricted energy spaces, a practice proven to be quite successful in many pure and disordered spin models~\cite{fytas08a,malakis04,malakis05,malakis06,fytas10}.  Estimating such ranges from a chosen pseudo-critical temperature one should be careful to account for the shift behavior of other important pseudo-critical temperatures and extend the subspace appropriately from both low- and high-energy sides in order to achieve an accurate estimation of all finite-size anomalies. At an initial stage of this work, preliminary comparative tests were also executed using the Metropolis algorithm~\cite{metropolis,landau_book} to provide  a benchmark, cf.~Fig.~\ref{fig:test}.

\begin{figure}
	\includegraphics[width=95mm]{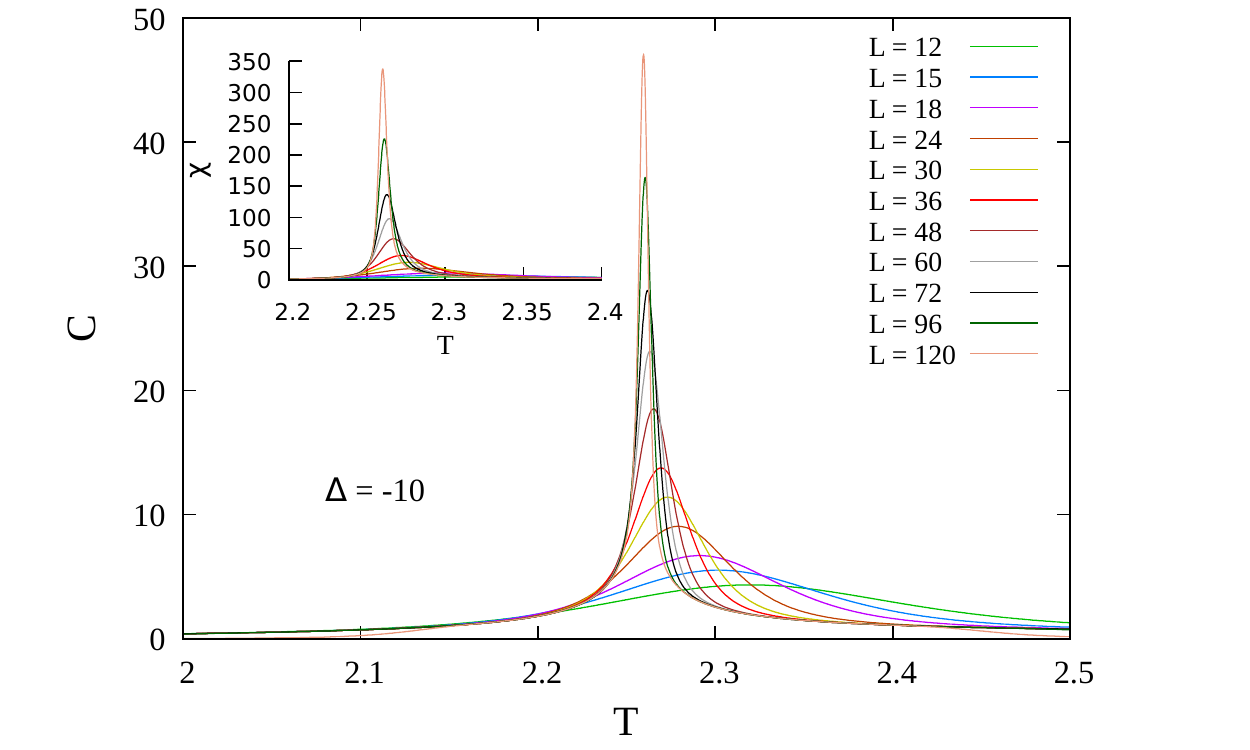}
	\caption{Specific-heat (main panel) and magnetic susceptibility (inset) curves corresponding to Eqs.~\eqref{eq:spec-heat} and \eqref{eq:susceptibility} at $\Delta = -10$ from Wang-Landau simulations. Upon increasing the system size the location of the peaks shifts to the left.
		\label{fig:curves_wl}}
\end{figure}

For the purposes of the present study we do not use the final estimate of $g(E)$ to compute thermodynamic averages but rather employ it as a weight function in a final production run. 
The sampled observables include estimates of the mean energy $\langle E \rangle$, the order parameter $\langle m\rangle$ which is estimated from  the root mean square average of the magnetization per site of the three sublattices A, B, and C~\cite{costa04b,costa16,jorge20}
\begin{equation}\label{eq:order-parameter}
	m = \sqrt{\frac{m_{\rm A}^{2} +m_{\rm B}^{2} +m_{\rm C}^{2}}{3}},
\end{equation}
the specific heat
\begin{equation}\label{eq:spec-heat}
	C = \left[\langle E^{2}\rangle - \langle E \rangle^{2}\right]/(NT^{2}),
\end{equation}
and the magnetic susceptibility 
\begin{equation}\label{eq:susceptibility}
 \chi = N\left[\langle m^{2}\rangle - \langle m\rangle^{2}\right]/T,
\end{equation}
where $N=L^{2}$ is the number of lattice sites. Characteristic specific-heat and magnetic susceptibility curves for the case $\Delta = -10$ obtained via Wang-Landau simulations are shown in Fig.~\ref{fig:curves_wl}.

\subsection{Multicanonical simulations}
\label{sec:muca}

\begin{figure}
	\includegraphics[width=95mm]{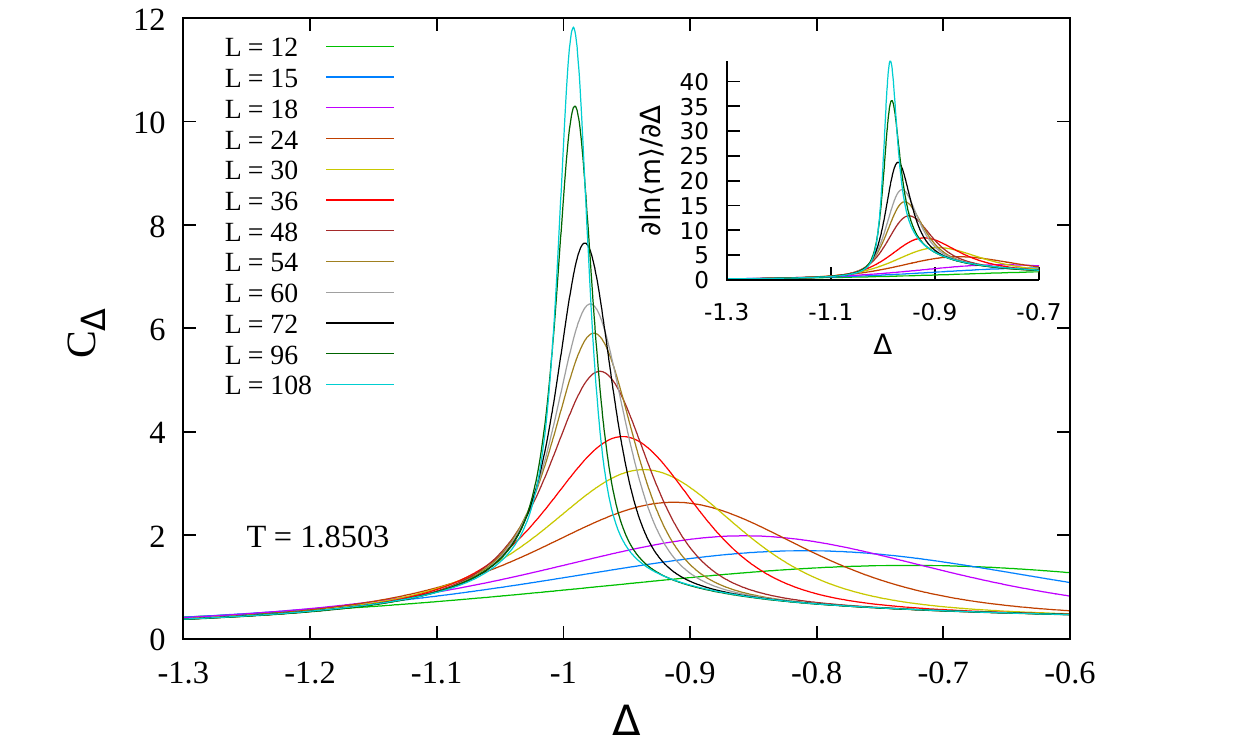}
	\caption{Specific-heat-like (main panel) and first-order logarithmic derivative of the order parameter (inset) curves obtained via multicanonical simulations at $T = 1.8503$. Similar to Fig.~\ref{fig:curves_wl} the location of the peaks shifts to the left as we increase the linear size of the system.
		\label{fig:curves_muca}}
\end{figure}

We now turn to the description of the multicanonical (MUCA) method~\cite{berg92}. In this approach, instead of using the canonical Boltzmann weight $e^{-\beta E}$, with $\beta = 1/T$, a correction function is introduced, designed to produce a flat histogram. For the purposes of the current work, the multicanonical method was applied with respect to the crystal-field energy $E_{\Delta}$ fixing the temperature and allowing us to continuously reweight at arbitrary values of $\Delta$ \cite{zierenberg15}. To this end, the partition function 
\begin{equation}\label{Z}
	\mathcal{Z} = \sum_{\{E_J, E_{\Delta}\}} g(E_J, E_{\Delta}) e^{-\beta(E_{J}+\Delta E_{\Delta})}
\end{equation}
is generalized to
\begin{equation}\label{Z_muca}
	\mathcal{Z} _\mathrm{MUCA} = \sum_{\{E_J, E_{\Delta}\}} g(E_J, E_{\Delta}) e^{-\beta E_{J}}~W\left(E_{\Delta}\right),
\end{equation}
where $g(E_J, E_\Delta)$ is the two-parametric density of states.
It follows that the equilibrium probability distribution in the multicanonical ensemble is
\begin{equation}\label{P_muca}
	P_\mathrm{MUCA}(E_J, E_\Delta)=\frac{g(E_J, E_{\Delta}) e^{-\beta E_{J}}W(E_{\Delta})}{\mathcal{Z} _\mathrm{MUCA}}.
\end{equation}
In order to produce a flat $E_{\Delta}$-histogram, by carrying out a summation with respect to $E_J$, the modified weight should be given by
\begin{equation}\label{W_muca}
	W(E_{\Delta}) \propto \mathcal{Z} _\mathrm{MUCA} \left[ \sum_{E_J} g(E_J, E_\Delta) e^{-\beta E_J}  \right]^{-1}.
\end{equation}

These weights can be calculated in an iterative fashion starting with an initial guess.
At the $n^\text{th}$ step spins are flipped using the weights $e^{-\beta E_{J}}W^{(n)}\left(E_\Delta\right)$ and the histogram $H^{(n)}(E_\Delta)$ of the energies $E_\Delta$ is sampled. After a specified number of spin-flip attempts the histogram is used to recalibrate the weights via $W^{(n+1)}\left(E_\Delta\right) = W^{(n)}\left(E_\Delta\right)/H^{(n)}(E_\Delta)$.
The process is completed when a sufficiently flat histogram has been achieved, after which a series of production runs is carried out. At each step the histogram $H^{(n)}(E_\Delta)$ satisfies the equation
\begin{widetext}
\begin{equation}\label{H_norm}
\langle H^{(n)}(E_\Delta)\rangle \propto P^{(n)}(E_\Delta) = \frac{1}{\mathcal{Z}_\mathrm{MUCA}}\sum_{E_J} g(E_J, E_{\Delta})e^{-\beta E_{J}} W^{(n)}(E_\Delta) \propto \frac{W^{(n)}(E_\Delta)}{W(E_\Delta)},
\end{equation}
\end{widetext}
justifying the scheme for updating the weights using sampled histograms.

We employ a parallel implementation of the multicanonical method~\cite{zierenberg13, gross18}, guided by its already successful application in the study of the Blume-Capel model in two and three dimensions~\cite{zierenberg15,zierenberg17,fytas18}. In this setup weights are distributed to parallel workers, each producing a histogram. At the end of each iteration all histograms are added into a single one which is then used to recalibrate the weights. Our simulations were implemented on an Nvidia K80 GPU, effectively running tens of thousands of simulation threads in parallel. Finally, the histogram flatness was tested using the Kullback-Leibler divergence~\cite{kullback51, gross18}. 

As the multicanonical method allows for continuously reweighting to any value of $\Delta$, canonical expectation values for an observable $O =O (\{\sigma\})$ at a fixed temperature can be obtained by estimating the expectation values
\begin{equation}
	\langle O\rangle_{\Delta}
	=
	\frac{\langle O(\{\sigma\}) e^{-\beta\Delta E_\Delta(\{\sigma\})}W^{-1}(E_\Delta) \rangle_{\rm MUCA}}
	{\langle   e^{-\beta \Delta E_\Delta(\{\sigma\}) }W^{-1}(E_\Delta) \rangle_{\rm MUCA}}.
	\label{eq:muca-reweight}
\end{equation}
In this framework, it is natural to compute $\Delta$-derivatives of observables rather than the usual $T$-ones. For instance, in place of the usual specific heat~(\ref{eq:spec-heat}) one may define a specific-heat-like quantity~\cite{zierenberg15}
\begin{equation}
	C_\Delta = \frac{1}{N} \frac{\partial E_J}{\partial\Delta}
	=
	-\left[ \left\langle E_J E_\Delta \right\rangle - \left\langle E_J \right\rangle \left\langle E_\Delta \right\rangle \right]/(NT),
\end{equation}
which shows the shift behavior expected from the usual specific heat as can be seen from the main panel of Fig.~\ref{fig:curves_muca}. Additionally, in order to obtain direct estimates of the critical exponent $\nu$ from finite-size scaling, one may compute the logarithmic derivatives of the order parameter~\cite{ferrenberg91,caparica00}
\begin{equation}
\label{eq:log_der}
	\frac{\partial\ln{\langle m^n \rangle}}{\partial \Delta}
	=
	-\left[\frac{\left\langle m^n E_\Delta \right\rangle}{\langle m^n \rangle} - \left\langle E_\Delta \right\rangle  \right]/T,
\end{equation}
see the inset of Fig.~\ref{fig:curves_muca} for the case $n=1$.

\begin{figure}
	\includegraphics[width=85mm]{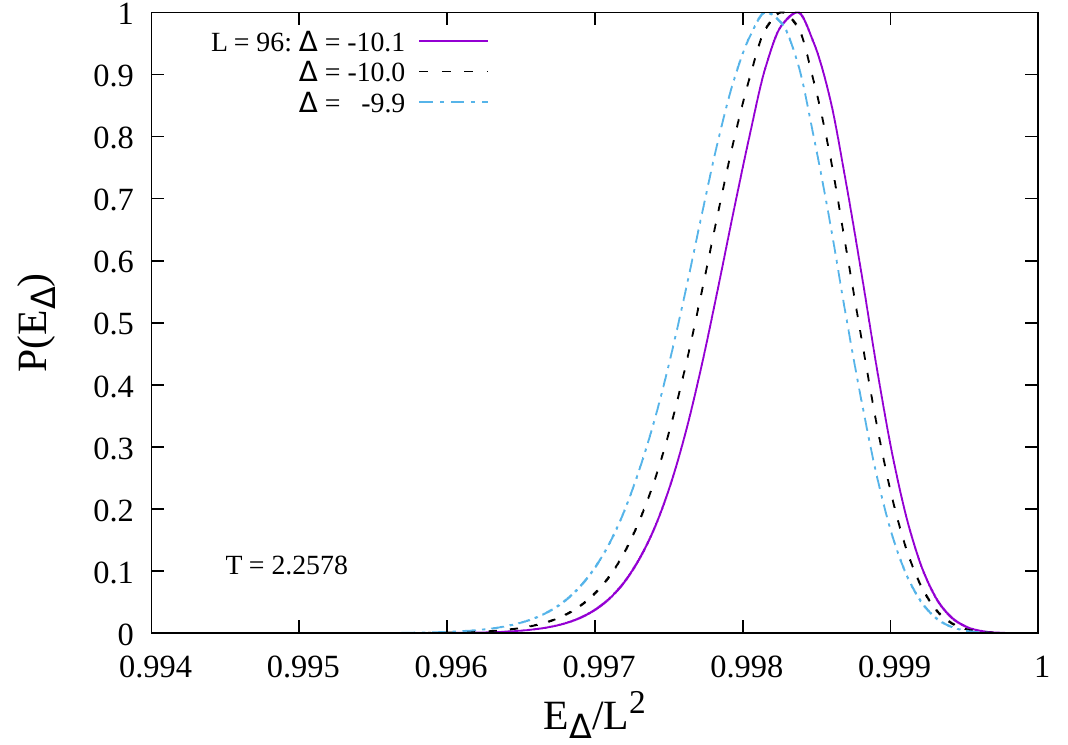}
	\caption{$P(E_{\Delta})$ for $L = 96$ obtained via multicanonical simulations at $T = 2.2578$. Results for three adjacent crystal-field values are shown.
		\label{fig:pdf_m10_MUCA}}
\end{figure}

Other useful observables accumulated during the multicanonical simulations are the magnetic susceptibility $\chi$ and the fourth-order Binder cumulant of the magnetization
\begin{equation}
\label{eq:binder_cum}
	U_m = 1 - \frac{\left\langle m^4 \right\rangle}{3\left\langle m^2 \right\rangle^2}.
\end{equation}

\subsection{Simulation parameters}
\label{sec:sim_parameters}

The numerical protocol described above was applied on triangular lattices with periodic boundary conditions. To accommodate not only the ferromagnetic ground state, but also the three ferrimagnetic ones, the allowed values of the linear size of the lattice $L$ must be a multiple of three~\cite{costa16}. In the course of our simulations we considered linear sizes within the range $12 \le L \le 120$ respecting this constraint. Wang-Landau simulations were carried out at two values of the crystal field, namely at $\Delta = -10$ and $-1$. We also performed a high-precision analysis using multicanonical simulations at the temperature $T=1.8503$ which roughly corresponds to the value $\Delta = -1$ of the phase diagram. Some additional simulations were conducted at $T = 2.2578$ (red dotted arrows in Fig.~\ref{fig:phase_diagram}). Finally, we would like to point out that for the fitting procedure discussed below in Sec.~\ref{sec:results} we restricted
ourselves to data with $L\geq L_{\rm min}$, adopting the standard $\chi^{2}$ test for goodness of the fit. Specifically, we considered a fit as being acceptable only if $10\% < Q < 90\%$, where $Q$ is the quality-of-fit parameter~\cite{press92}. 

\section{Results }
\label{sec:results}

\subsection{Order of the transition}
\label{sec:results_a}

As discussed above, there have been recent reports of first-order transition features even along the putatively continuous part of the transition line~\cite{jorge21,jorge20}. In particular, the authors of Ref.~\cite{jorge21} using Wang-Landau simulations and a system with linear size $L = 16$ studied the shape of the energy probability distribution, $P(E)$, at several values of the crystal field, $\Delta = \{-2, 0, 1, 1.5\}$. Indeed they observed that $P(E)$ exhibits two peaks of the same height close to the estimated transition temperature (see Fig. 2 in Ref.~\cite{jorge21}). It is well known that a double-peak structure in the density function in finite systems is an expected precursor of the two $\delta$-peak behavior in the thermodynamic limit occurring for a first-order phase transition~\cite{binder84,binder87}.  

\begin{figure}[htbp]
	\includegraphics[width=85mm]{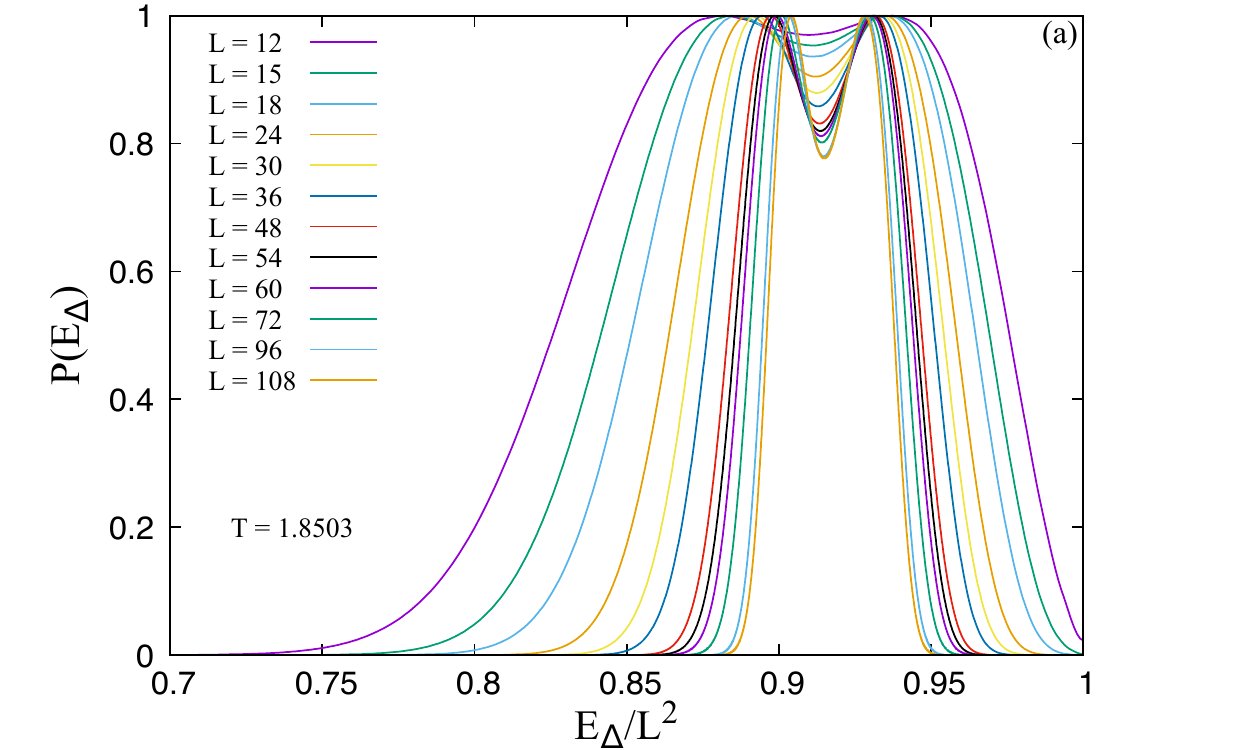}\\
	\includegraphics[width=85mm]{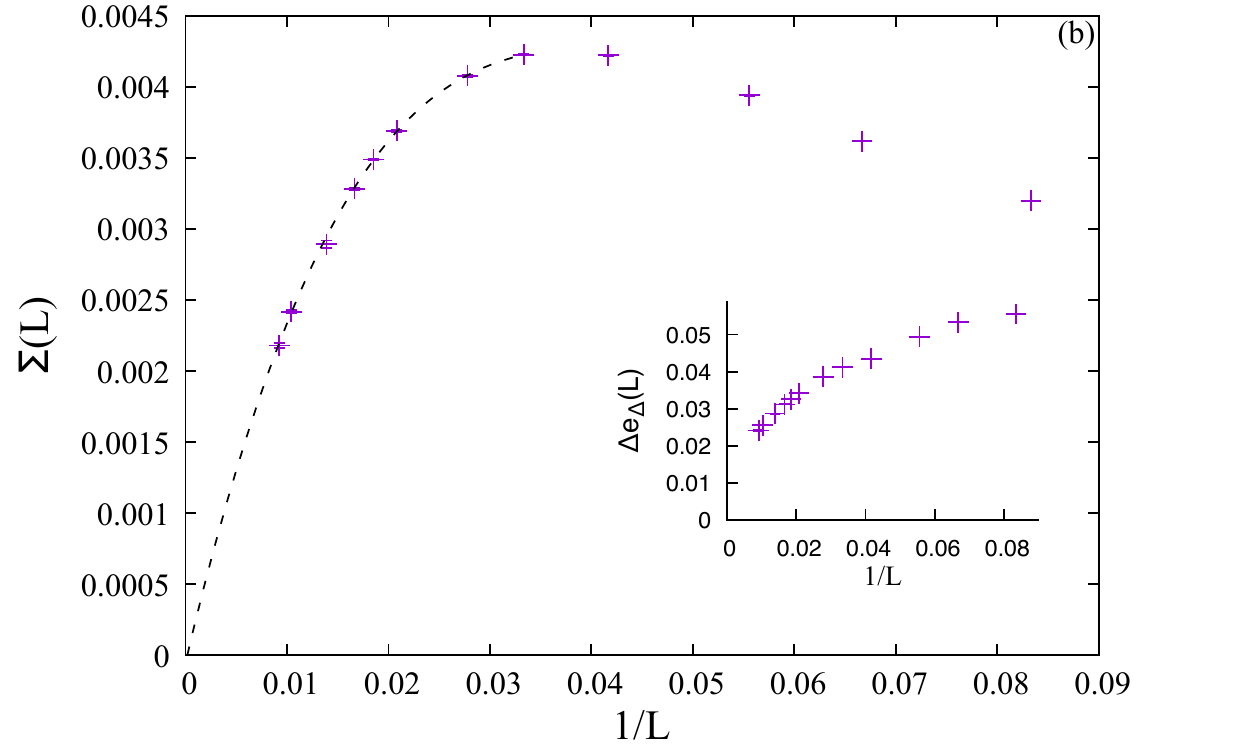}
	\caption{\label{fig:first_order_muca}
		(a) Reweighted probability density functions $P(E_{\Delta})$ for various system sizes. Upon increasing $L$ the distance between the two peaks is decreasing. (b) Limiting behavior of the corresponding surface tension $\Sigma(L)$ (main panel) and latent heat $\Delta e_{\Delta}(L)$ (inset). Results obtained via multicanonical simulations at $T = 1.8503$.}
\end{figure}

In order to provide clarity regarding the transition order, we studied the reweighted probability density function $P(E_{\Delta})$ normalized to unity as obtained directly from the multicanonical simulations. This approach has already been successfully applied to a number of models undergoing first-order phase transitions~\cite{zierenberg15,zierenberg17,fytas18}. We start with Fig.~\ref{fig:pdf_m10_MUCA} which illustrates the probability density function $P(E_{\Delta})$ for a system with linear size $L = 96$ at the temperature $T = 2.2578$ corresponding to $\Delta = -10$ (see Fig.~\ref{fig:phase_diagram}). Clearly, no sign of a double-peak structure is observed which would indicate the presence of a first-order transition. On the other hand, as we lower the temperature gradually to $T=1.8503$ (corresponding to $\Delta = -1$), first-order-like characteristics appear -- see Fig.~\ref{fig:first_order_muca}(a) -- in agreement with the results of Ref.~\cite{jorge20} for the case $\Delta = 0$. 

This observation calls for  a systematic analysis of the relevant surface tension and latent heat of the transition as suggested by Lee and Kosterlitz~\cite{lee90}. In fact, the multicanonical method is instrumental for this purpose as it allows the direct estimation of the barrier associated with the suppression of states during a first-order phase transition. Considering distributions with two peaks of equal height (eqh)~\cite{borgs92}, such as the ones shown in Fig.~\ref{fig:first_order_muca}(a), allows one to extract the free-energy like barrier in the $E_{\Delta}$-space,
\begin{equation}
	\Delta F(L) = \frac{1}{2\beta\Delta}\ln\left(\frac{P_{\rm
			max}}{P_{\rm min}}\right)_{\rm eqh},
\end{equation}
where $P_{\rm max}$ and $P_{\rm min}$ are the maximum and
local minimum of the distribution $P(E_{\Delta})$, respectively. The
resulting barrier connects a spin-$0$ dominated regime ($E_{\Delta}$
small) and a spin-$\pm1$ rich phase ($E_{\Delta}$ large). The corresponding surface tension $\Sigma(L) = \Delta F(L)/L$ is expected to scale as $\Sigma(L)  = \Sigma_{\infty} + c_1L^{-1} + \mathcal{O}\left(L^{-2}\right)$ in two dimensions, possibly with higher-order corrections~\cite{Nussbaumer2006,Nussbaumer2008,Bittner2009}. Similarly we may define the latent heat of the transition $\Delta e_{\Delta}(L)$, where $e_{\Delta} = E_{\Delta}/L^2$, as the difference in energies of the two peaks. The scaling behavior of these observables is presented in Fig.~\ref{fig:first_order_muca}(b). Note the existence of a crossover length $L^{\ast} \approx 30$ where the slope in the trend of $\Sigma(L)$ changes sign indicating strong finite-size effects. The dashed line in the main panel shows a fit including third-order corrections terms for $L\geq L^{\ast}$ giving a practically zero value of $\Sigma_{\infty} = -5\times 10^{-5} \pm 11\times 10^{-5}$. A similar, but somehow slower downward trend is also observed in the latent heat presented in the inset of Fig.~\ref{fig:first_order_muca}(b). 

Thus, our numerical data and analysis highlight the presence of non-negligible finite-size effects that become more pronounced while arriving at the pentacritical point, and that could possibly account for misleading previous conclusions that the transition is of first order. However, we should note that for the present spin-$1$ Baxter-Wu model reaching  an unquestionable conclusion is a very difficult numerical exercise, that is also heavily undermined by the ambiguity in the location of its pentacritical point~\cite{dias17,jorge21}. 

\begin{figure}
	\includegraphics[width=95mm]{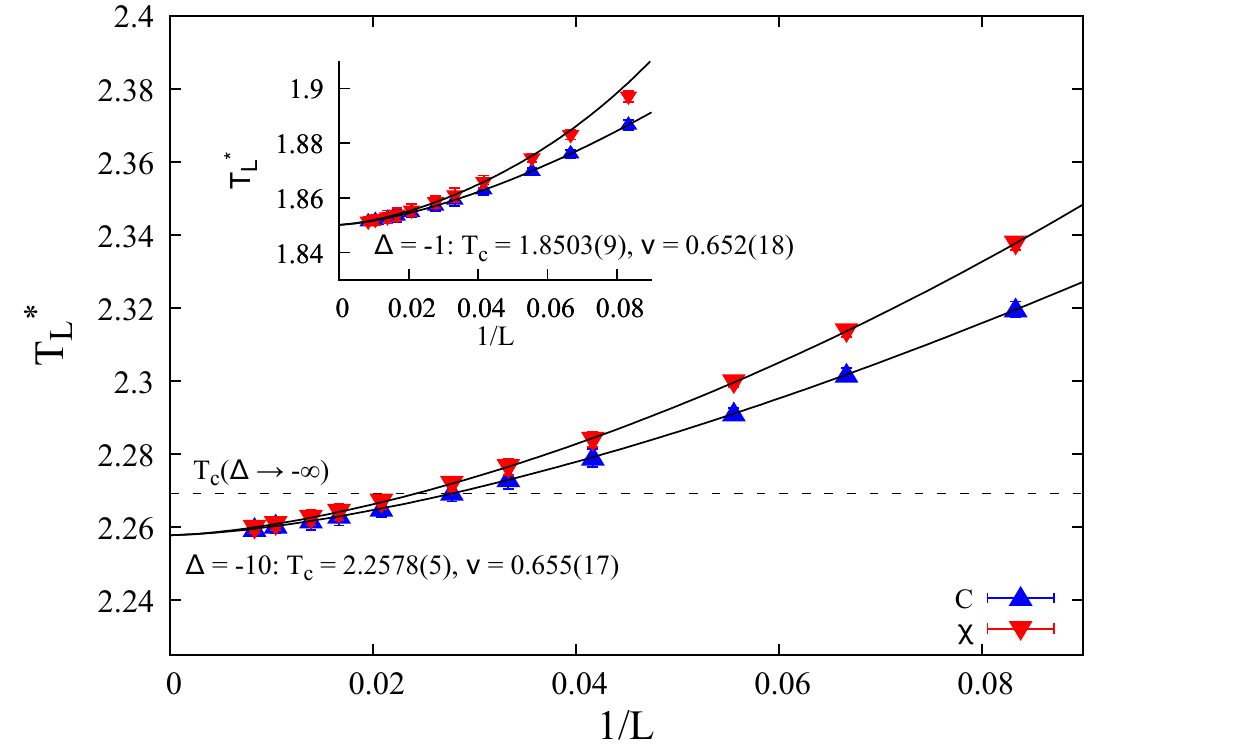}
	\caption{Shift behavior of the peak locations of the specific heat and magnetic susceptibility as a function of the inverse linear system size at $\Delta = -10$ (main panel) and $\Delta = -1$ (inset). The black dashed line denotes the critical temperature of the model at $\Delta \rightarrow - \infty$, \emph{i.e.}, the critical temperature of the spin-$1/2$ Baxter-Wu model. In both panels the black solid lines are joint fits of the form~(\ref{eq:pseudo_T_scaling}). Data produced with the Wang-Landau algorithm.
		\label{fig:pseudocritical_WL}}
\end{figure}

\subsection{Finite-size scaling and universality}
\label{sec:results_b}

Having established the continuous nature of the transition we proceed to a detailed finite-size scaling analysis of the numerical data designed to probe the universality class of the second-order transition. In what follows we show a selection of results obtained via Wang-Landau and multicanonical simulations for a range of observables that support the original expectation that the spin-$1$ Baxter-Wu model in a crystal field belongs to the universality class of the $4$-state Potts model.

\begin{figure}[htbp]
	\includegraphics[width=85mm]{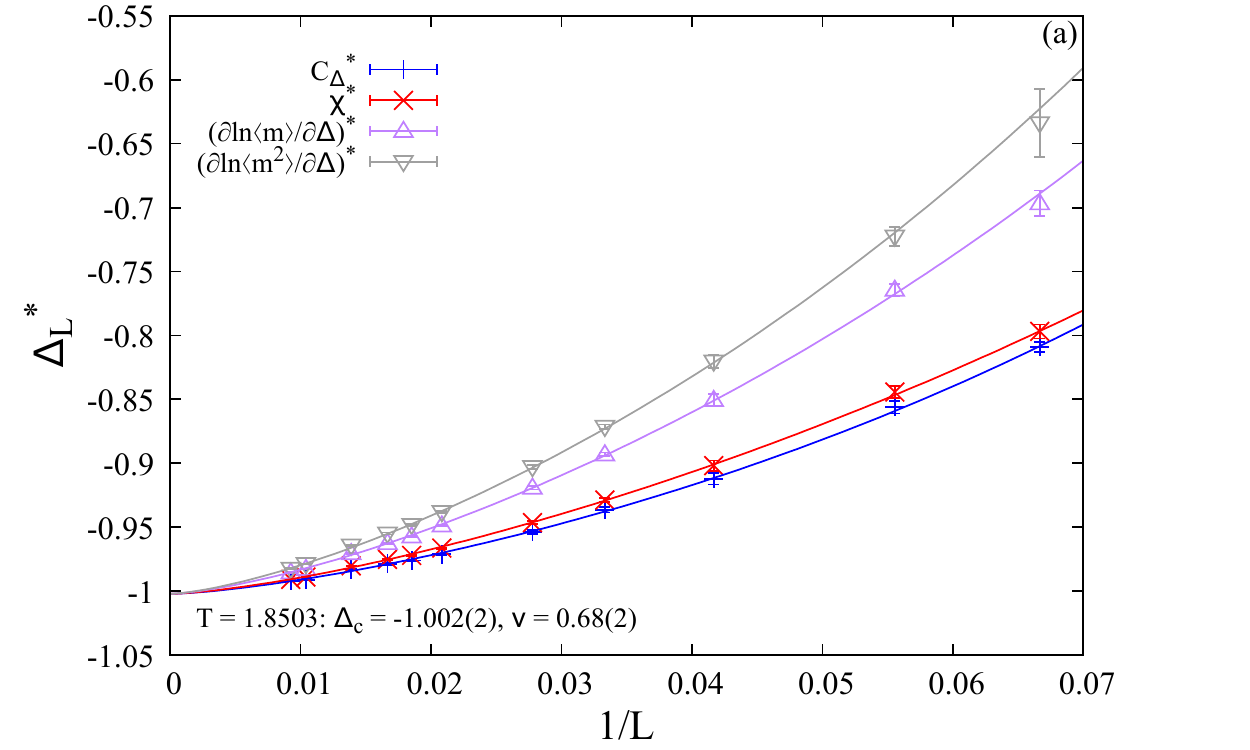}\\
	\includegraphics[width=85mm]{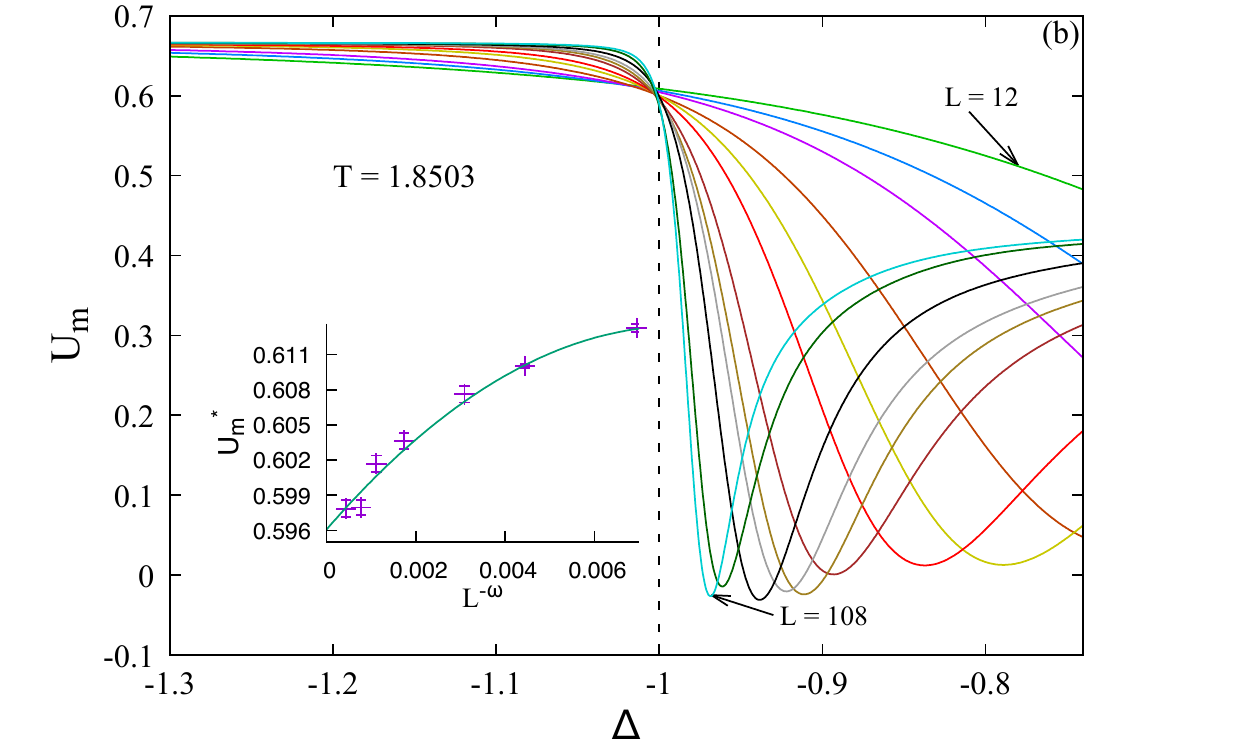}
	\caption{\label{fig:pseudocritical_muca}
		(a) Shift behavior of several pseudocritical fields as a function of the inverse linear system size. (b) Fourth-order Binder cumulant curves of the order parameter. The black vertical dashed line marks the value $\Delta = -1$. The inset shows the limiting behavior of the crossings $U_{m}^{\ast}$ on pairs of lattice sizes $(L, 2L)$. Data produced at $T = 1.8503$ via multicanonical simulations.}
\end{figure}

In order to extract critical temperatures $T_{\rm c}(\Delta)$ and crystal fields $\Delta_{\rm c}(T)$ of the system as well as a first estimate of the correlation-length exponent $\nu$ we present in Fig.~\ref{fig:pseudocritical_WL} the shift behavior of suitable pseudocritical temperatures, $T_{L}^{\ast}$, defined as the peak locations of the specific-heat $C$ and susceptibility $\chi$ curves of Fig.~\ref{fig:curves_wl}. Two data sets are shown, corresponding to $\Delta = -10$ (main panel) and $\Delta = -1$ (inset). For each value of $\Delta$ the solid lines are joint fits of the expected power-law behavior \begin{equation}\label{eq:pseudo_T_scaling}
	T^{\ast}_{L}=T_{\rm c}+bL^{-1/\nu}(1+b'L^{-\omega})
\end{equation}
to the data, where the correction-to-scaling exponent $\omega$ is fixed hereafter to the accepted value $2$~\cite{alcaraz97,alcaraz99,dias17,costa16}. Using $L_{\rm min} = 12$ we obtain the values $T_{\rm c} (\Delta = -10)= 2.2578(5)$ and $T_{\rm c}(\Delta = -1) = 1.8503(9)$ in excellent agreement with the values $2.2578(116)$ and $1.8503(94)$, respectively, reported in Ref.~\cite{dias17} using conventional finite-size scaling. More importantly, our estimates $\nu = 0.655(17)$ for $\Delta = -10$ and $\nu = 0.652(18)$ for $\Delta = -1$ agree nicely with the value $\nu=2/3$ of the $q=4$ Potts universality class. 

Similarly, in Fig.~\ref{fig:pseudocritical_muca}(a) we present the shift behavior of several pseudocritical fields, $\Delta_{L}^{\ast}$, defined as the peak locations of the $\Delta$-dependent curves defined in Sec.~\ref{sec:methods}. A simultaneous fit of the form
\begin{equation}\label{eq:pseudo_Delta_scaling}
	\Delta^{\ast}_{L}=\Delta_{\rm c}+bL^{-1/\nu}(1+b'L^{-\omega}),
\end{equation}
using $L_{\rm min} = 15$ provides the estimates $\Delta_{\rm c} (T = 1.8503) = -1.002(2)$ and $\nu = 0.68(2)$ in very good agreement with the results of Fig.~\ref{fig:pseudocritical_WL}. Moreover, in the main plot of Fig.~\ref{fig:pseudocritical_muca}(b) typical curves of the fourth-order Binder cumulant $U_m$~(\ref{eq:log_der}) are shown, where the location of crossing point also agrees nicely with the value $\Delta = -1$ (see also Figs.~\ref{fig:phase_diagram} and \ref{fig:pseudocritical_WL}). 

\begin{figure}
	\includegraphics[width=95mm]{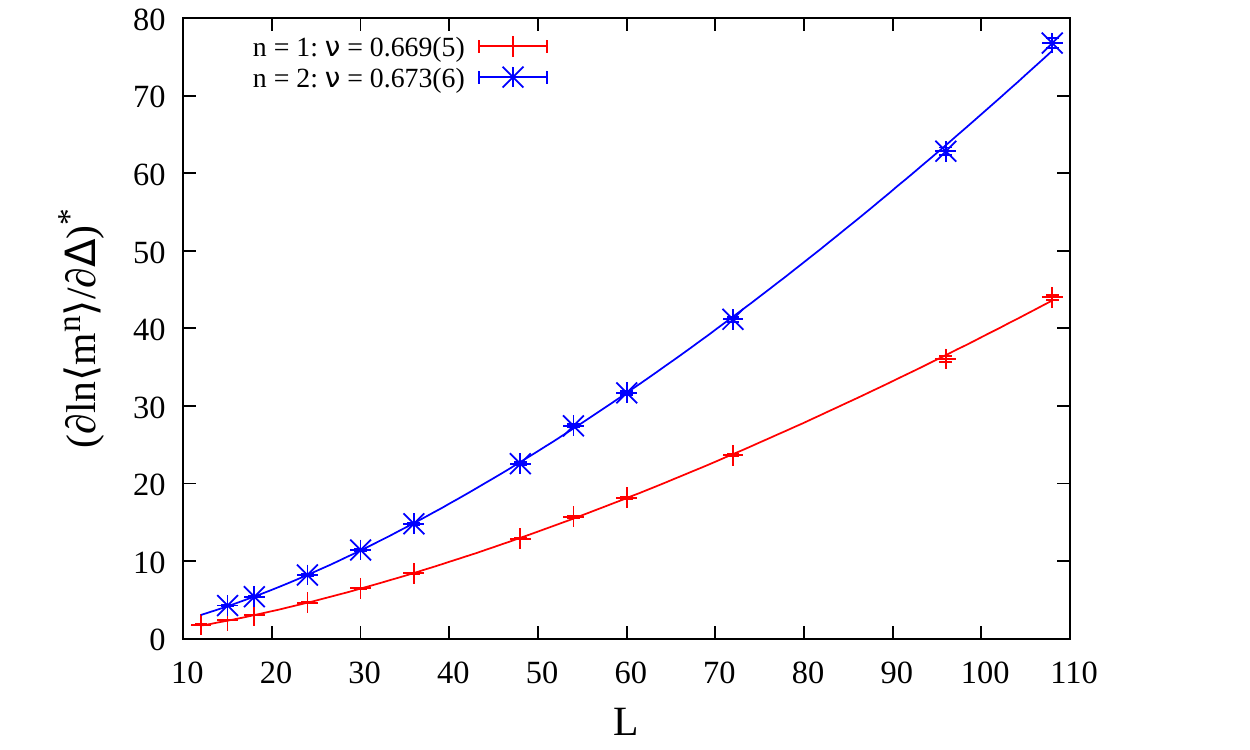}
	\caption{Finite-size scaling of the logarithmic derivatives of powers $n=1$ and $2$ of the order parameter at $T = 1.8503$. The solid lines are fits of the form~(\ref{eq:log_der_scaling}). Results obtained via multicanonical simulations. 
		\label{fig:scaling_derivatives_muca}}
\end{figure}

Additional estimates for the critical exponent $\nu$ can be obtained via the scaling of the maxima of the logarithmic derivatives of the order parameter~(\ref{eq:log_der}). Since these are dimensionless quantities, we expect them to scale as
\begin{equation}\label{eq:log_der_scaling}
	\left(\frac{\partial\ln{\langle m^n \rangle}}{\partial \Delta}\right)^{\ast} \sim L^{1/\nu}(1+b'L^{-\omega}).
\end{equation}
The numerical data for $n=1$ and $n=2$ obtained from multicanonical simulations at $T = 1.8503$ are shown in Fig.~\ref{fig:scaling_derivatives_muca}, and the solid lines are power-law fits of the form~(\ref{eq:log_der_scaling}) with $L_{\rm min} = 18$ giving $\nu = 0.669(5)$ and $0.673(6)$, respectively. Again these results point to the expected $2/3$ value of the $q=4$ Potts universality class. 

\begin{figure}[htbp]
	\includegraphics[width=85mm]{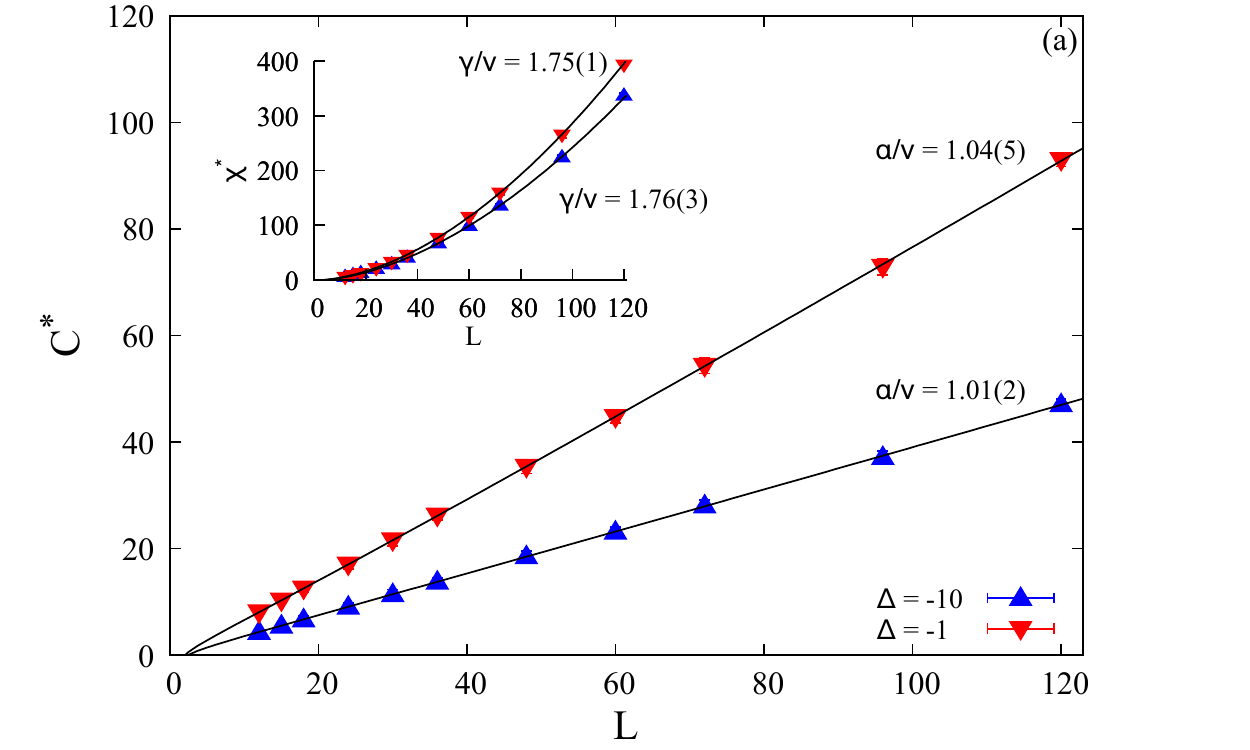}\\
	\includegraphics[width=85mm]{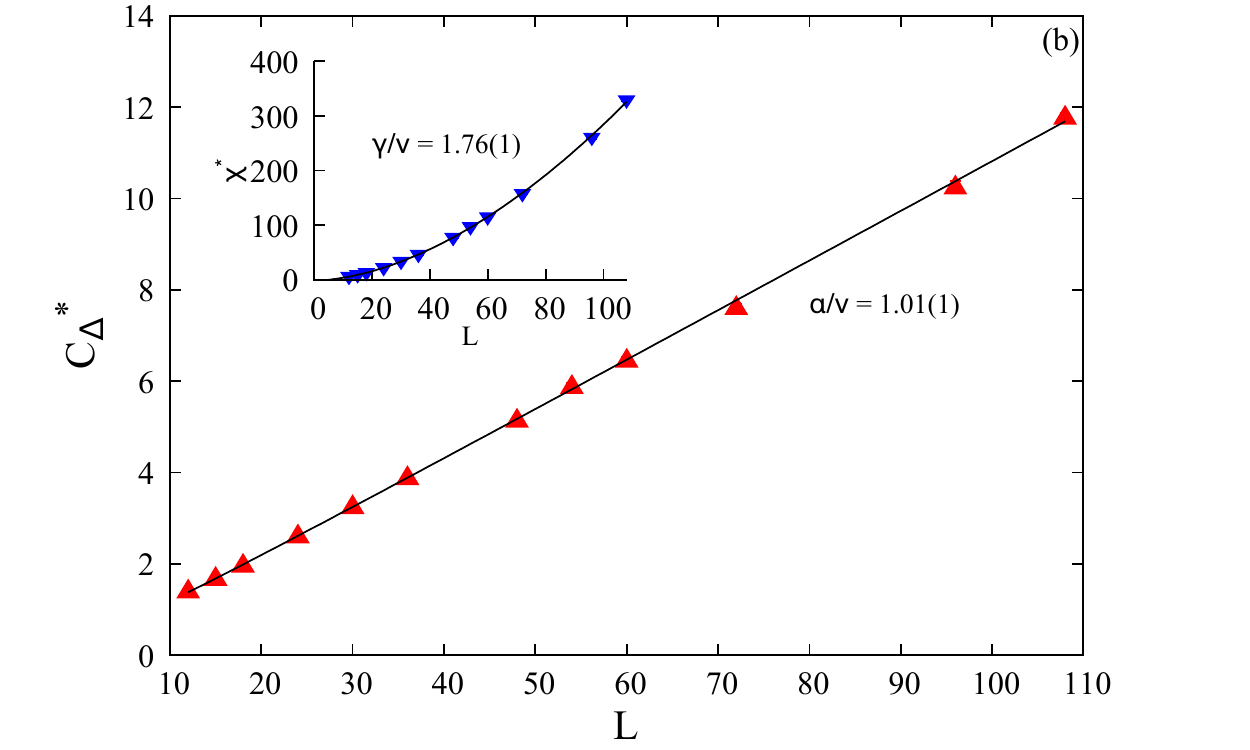}
	\caption{\label{fig:scaling}
		(a) Finite-size scaling behavior of $C^{\ast}$ (main panel) and $\chi^{\ast}$ (inset) at $\Delta = -10$ and $\Delta = -1$. Data produced with the Wang-Landau algorithm. (b) Similar analysis of data produced at $T = 1.8503$ via multicanonical simulations.}
\end{figure}

We now turn to the finite-size scaling behavior of the maxima of the specific heat ($C^{\ast}$ and $C^{\ast}{(\Delta)}$, respectively) and magnetic susceptibility ($\chi^{\ast}$) in order to probe the critical exponent-ratios $\alpha/\nu$ and $\gamma / \nu$, respectively. Figure~\ref{fig:scaling} presents numerical data obtained via the Wang-Landau algorithm [panel (a) at $\Delta = -10$ and $-1$] and the multicanonical approach [panel (b) at $T = 1.8503$]. In all cases the solid lines are fits of the form
\begin{equation}\label{eq:spec-heat_scaling}
	C^{\ast}_{(\Delta)} \sim L^{\alpha/\nu}(1+b'L^{-\omega})
\end{equation}
and
\begin{equation}\label{eq:susceptibility_scaling}
	\chi^{\ast} \sim L^{\gamma/\nu}(1+b'L^{-\omega}),
\end{equation}
choosing $L_{\rm min} = 18$. The obtained estimates of $\alpha/\nu$ and $\gamma/\nu $ are listed in the panels (see also Tab.~\ref{tab:summary} below) and are clearly compatible to the exact values $\alpha/\nu = 1$ and $\gamma/\nu = 7/4$ of the $4$-state Potts universality class~\cite{baxter73}. As a side note, error propagation and $\nu$ values from Fig.~\ref{fig:pseudocritical_WL} suggest that $\alpha = 0.662(22)$ and $0.678(38)$ for $\Delta = -10$ and $\Delta = -1$, respectively~\footnote{We should point out that the indicated error in $\alpha$ is not fully reliable due to the statistical correlation of the estimates of $\nu$ and the ratio $\alpha/\nu$~\cite{weigel09}.}. 

At this point we would like to make a remark on the additional correction term $b'L^{-\omega}$ appearing in the fits of Figs.~\ref{fig:pseudocritical_WL}--\ref{fig:scaling}. Although in the work of Jorge \emph{et al.}~\cite{jorge16} for the spin-$1/2$ model, critical exponents were obtained with very good accuracy and without the need for corrections to scaling, the situation here is rather different. In particular the values of scaling amplitudes $b$ and $b'$ in Eqs.~(\ref{eq:pseudo_T_scaling})--(\ref{eq:susceptibility_scaling}) are comparable and in particular the values of $b'$ fluctuate within the range $1 - 20$ for the various observables and cannot be neglected. Additionally, from our overall comparative tests we may safely conclude that the fitting quality measured in terms of the probability $Q$ is indeed improved when the correction term $b'L^{-\omega}$ is included.

\begin{figure}[tb]
	\includegraphics[width=95mm]{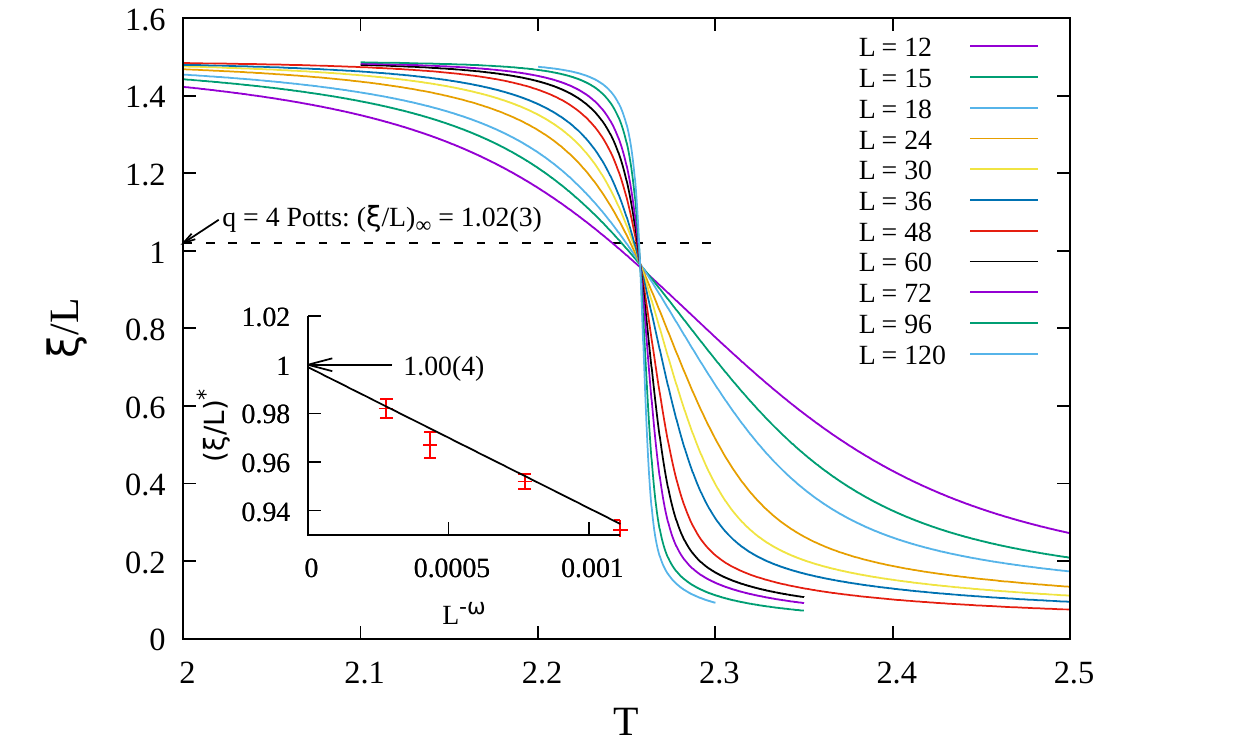}
	\caption{{\bf Main panel}: Typical $\xi/L$ curves as a function of the temperature obtained from Wang-Landau simulations for all pairs of system sizes studied and for $\Delta = -10$. The temperature area of the crossings conforms to the value $T_{\rm c} = 2.2578$ of Fig.~\ref{fig:pseudocritical_WL}. {\bf Inset}: Finite-size scaling of the correlation-length ratios at their crossing points, $(\xi/L)^{\ast}$. Results are shown for the largest pairs $(L,2L)$ of system sizes: $(30, 60)$, $(36,72)$, $(48, 96)$, and $(60, 120)$. The solid line shows a linear in $L^{-\omega}$ extrapolation to $L \rightarrow \infty$. The black dashed line marks the value of $(\xi/L)_{\infty}$ of the $4$-state Potts model, as taken from Ref.~\cite{salas97}.  
		\label{fig:cor_length}}
\end{figure}

Universality classes are characterized by a whole range of universal quantities, which include critical exponents but also certain amplitude ratios $g$~\cite{shchur10,fytas18,fytasRFIM,hasenbusch10}. In contrast to exponents, amplitude ratios depend on additional system properties, such as the lattice geometry and boundary conditions. In the present work we study two of these universal amplitudes, namely the well-known Binder cumulant $g = U_{m}$, see Eq.~(\ref{eq:binder_cum}), and the ratio of the correlation length over the linear system size, $g = \xi/L$; typical curves of $\xi/L$ at $\Delta = -10$ are shown in the main panel of Fig.~\ref{fig:cor_length}. For the estimation of $\xi$ we used the well-known second-moment definition~\cite{cooper1989, ballesteros2001,fytas18}: From the Fourier transform of the spin field, $\hat{\sigma}(\mathbf{k}) = \sum_{\bf x}\sigma_{\bf x}\exp(i{\bf kx})$, we determined
\begin{equation}
\begin{split}
 F =& \left\langle
|\hat{\sigma}(2\pi/L,0)|^2+|\hat{\sigma}(0,2\pi/L)|^2\right.\\
&+\left.|\hat{\sigma}(2\pi/L,2\pi/L)|^2\right\rangle/3
\end{split}
\end{equation}
and attained the correlation length via~\cite{ballesteros2001}
\begin{equation}
    \xi  \equiv  \frac{1}{2\sin(\pi/L)}\sqrt{\frac{\langle m^2\rangle}{F}-1}
\end{equation}
To monitor the size evolution and limiting behavior of these amplitudes we employ the quotients method~\cite{fytasRFIM,night,bal96}: The crystal field (resp.\ temperature) where
$g_{2L} / g_{L} = 2$, i.e., where the curves of $U_{m}$ (resp.\ $\xi/L$) of the sizes
$L$ and $2L$ cross, defines the finite-size pseudocritical points (see Fig.~\ref{fig:pseudocritical_muca}(b) and also Fig.~\ref{fig:cor_length}). Let us denote the value of $g$ at these crossing points
as $g^{\ast}$. Within the framework of the quotients method a scaling of the form 
$g^{\ast} = g_{\infty}+\mathcal{O}(L^{-\omega})$
is expected, where $g_{\infty}$ is a universal value.

In the inset of Fig.~\ref{fig:pseudocritical_muca}(b) we provide an estimate of the universal Binder cumulant $U_{m,\infty}$ extracted from this sequence $U_m^\ast$. The solid line is a second-order polynomial fit in $L^{-\omega}$, yielding $U_{m,\infty} = 0.596(6)$ in very good agreement with the graphical estimate $0.595$ obtained by Capponi \emph{et al.}~\cite{capponi14}. Similarly, in the inset of Fig.~\ref{fig:cor_length} we show the infinite-size extrapolation of $(\xi/L)^{\ast}$ for the four largest pairs of system sizes as listed in the caption of this figure. The solid line is a linear fit in $L^{-\omega}$, leading to
\begin{equation}\label{eq:ratio_high_T}
	\left(\frac{\xi}{L}\right)_{\infty, \; \rm spin-1\; BW} = 1.00(4).
\end{equation}
We recall the value of $(\xi / L)_{\infty}$ for the two-dimensional $q=4$ Potts model with periodic boundary conditions from the seminal work of Salas and Sokal~\cite{salas97}
\begin{equation} \label{eq:ratio-exact}
	\left(\frac{\xi}{L}\right)_{\infty, \; q=4 \; \rm Potts} = 1.02(3).
\end{equation}
A comparison of the results of Eqs.~(\ref{eq:ratio_high_T}) and (\ref{eq:ratio-exact}) consists our final universality check which succeeds within $\sim 2\%$ accuracy.
We note here that an alternative approach that allows to fit the whole set of data points to a two-parameter finite-size scaling ansatz that includes the temperature can be found in Ref.~\cite{salas20}.

\begin{table*}
	\caption{\label{tab:summary} An overview of exact and numerical results for the $4$-state Potts model and the spin-$1/2$ Baxter-Wu model, together with a summary of numerical results for the spin-$1$ Baxter-Wu model in a crystal field obtained in the current work via: (i) Wang-Landau simulations at fixed values of the crystal field $\Delta$ (columns 4 and 5) and (ii) multicanonical simulations at a fixed temperature $T$ (column 6).}
	\begin{ruledtabular}
		\begin{tabular}{lccccc}
			& $4$-state Potts & spin-$1/2$ Baxter-Wu  & & spin-$1$ Baxter-Wu  \\ \hline \hline
			& Ref.~\cite{domany78} &  Ref.~\cite{baxter73} & $\Delta = -10$ & $\Delta = -1$ & $T = 1.8503$\\
			\hline
			$\nu$                  &  $2/3$  & $2/3$ &   $0.655(17)$     &    $0.652(18)$ & $0.671(6)$\footnote{This estimate corresponds to the average value of $\nu$ obtained from the fits of Fig.~\ref{fig:scaling_derivatives_muca}. Cross-correlations were not taken into account, but see Ref.~\cite{weigel09}.} \\
			$\alpha/\nu$           &  $1$    & $1$ &   $1.01(2)$     &   $1.04(5)$  & $1.01(1)$ \\
			$\gamma/\nu$           &  $7/4$   & $7/4$  &   $1.76(3)$   &  $1.75(1)$    & $1.76(1)$  \\
			\hline \hline
			$(\xi/L)_{\infty}$                &  $1.02(3)$~\cite{salas97} &  & $1.00(4)$     &    &    \\
			\hline \hline
			$U_{m,{\infty}}$                &   & $\sim 0.595$~\cite{capponi14} &      &    &  $0.596(6)$  \\
			\hline \hline
			$T_{\rm c}(\Delta)$ or $\Delta_{\rm c}(T)$      &    &    &   $2.2578(5)$      &  $1.8503(9)$   &  $-1.002(2)$ \\
		\end{tabular}
	\end{ruledtabular}
\end{table*}

\section{Summary and Outlook}
\label{sec:summary}

We presented here an extensive numerical study of scaling and universality in the phase diagram of the dilute Baxter-Wu model. Using a highly optimized combination of Wang-Landau simulations that cross the transition at constant crystal field $\Delta$ and multicanonical simulations operating at fixed temperature $T$, we covered a range of the transition line defined by $\Delta \le -1$. We provided strong evidence for a continuous nature of the transition in this regime. The previously reported first-order signature of the transition on approaching the pentacritical point is also seen here but a careful finite-size scaling analysis shows that they are a finite-size effect with a crossover-length $L^\ast \approx 30$ beyond which the first-order character disappears, at least for the region of interest in this work. It would be instrumental to probe in detail the system's behavior at positive crystal-field values and in particular within the regime $0.89  \lesssim \Delta \lesssim 1.68$ as marked by the two $\Delta_{\rm pp}$-estimates of Refs.~\cite{dias17,jorge21}, where the most strong first-order characteristics of the transition have been recorded~\cite{jorge21,jorge20}. Everywhere in the second-order regime our analysis clearly shows consistency with the universality class of the $4$-state Potts model. From the accuracy in the determination of critical exponents one may conclude that logarithmic corrections to scaling are indeed absent in this model as compared to the $4$-state Potts model. On the other hand, including the expected correction-to-scaling term $\mathcal{O}(L^{-\omega})$, with $\omega = 2$, at first-order, is necessary to achieve the optimum merit of the fits. A comparative overview of our results is provided in Tab.~\ref{tab:summary}. While it is clear from our results that strong scaling corrections appear as the pentacritical point where the transition changes to first-order is approached, the exact location of this pentacritical point and its universality class were not considered here. This question is left for future work. To conclude, we hope that this work settles some of the previously reported controversies over the critical behavior of the spin-$1$ Baxter-Wu model and lays the foundation for intriguing extensions. One such interesting line of research would be to unveil the effect of quenched disorder in both parts of the phase diagram of the model. 

\begin{acknowledgments}
We would like to thank the two anonymous referees who have helped us improve our manuscript with their instructive comments. We are also grateful to Jes\'us Salas and Joao Antonio Plascak for their fruitful communication on the problem. We acknowledge the provision of computing time on the parallel computer cluster \emph{Zeus} of Coventry University and T\"{U}B\.{I}TAK ULAKB\.{I}M (Turkish agency), High 
Performance and Grid Computing Center (TRUBA Resources). 
\end{acknowledgments}

{}

\begin{thebibliography}{199}
\bibitem{comment1} Note that the value $\Delta = 1$ is now known to correspond to the first-order regime of the phase diagram, see Ref.~\cite{dias17}.
\end{thebibliography}
\end{document}